\documentclass[epj]{svjour}
\usepackage{graphics}
\newcommand{\vecr}{{\bf r}}
\begin{document}
\title{Giant resonances in the deformed continuum}
\author{Takashi Nakatsukasa\inst{1} \and Kazuhiro Yabana\inst{2}% etc
}                     % Do not remove
\offprints{}          % Insert a name or remove this line
\institute{Physics Department, Tohoku University, Sendai 980-8578, Japan
 \and Institute of Physics, University of Tsukuba, Tsukuba 305-8571, Japan}
\date{Received: date / Revised version: date}
\abstract{
Giant resonances in the continuum for deformed nuclei are studied with
the time-dependent Hartree-Fock (TDHF) theory
in real time and real space.
The continuum effect is effectively
taken into account by introducing a complex Absorbing Boundary Condition (ABC).
} %end of abstract

 % end of PACS codes
\maketitle

Giant resonances (GR) are the most prominent feature of nuclei at
high excitation energy.
Recently, the GR becomes an active issue again in relation with the structure
of unstable nuclei.
The major part of GR's strength is embedded in the continuous energy spectrum,
at energies higher than the neutron and proton separation energies.
Thus, the coupling to the continuum is a key issue to understand
the excitation mechanism and damping of the GR.

It has been well-known that
the effects of the continuum can be properly treated in the random-phase
approximation (RPA) with Green's function in the coordinate space \cite{SB75}.
However, it is very difficult to directly apply the method to deformed nuclei
because construction of the Green's function becomes a difficult task
for the multi-dimensional space.
Recently, we have developed a numerical method of constructing the
three-dimensional (3D) Green's function with a proper boundary condition
and studied photoabsorptions of molecules \cite{NY01-1}.
Although the method works successfully in calculations of response functions
in electron systems, complexity of nuclear Hamiltonian makes its application
to nuclear physics harder \cite{NY01-2}.
Therefore, we investigate an alternative treatment of the continuum in
deformed systems; {\it Absorbing Boundary Condition (ABC)} method.
The ABC method was extensively utilized in a field of the quantum chemistry,
in order to study chemical reactions (for instance, see Ref.~\cite{SM92}).
We have shown that the ABC method is also very useful to treat
the electronic continuum in molecules and clusters \cite{NY01-1}.
We are also applying the ABC method to
studies of the continuum effects in nuclear response \cite{NY02-1,NY02-2}
and in nuclear reaction \cite{UYN02-1,UYN02-2}.

The essential trick for the treatment of the continuum in the ABC method is 
to allow the infinitesimal imaginary part in the Green's function,
$i\epsilon$, to be a function of coordinate and finite, $i\epsilon(\vecr)$
\cite{SM92}.
The $\epsilon(\vecr)$ should be zero in the interaction region and
positive outside the physically relevant region of space.
This is equivalent to adding the absorbing potential
to the original Hamiltonian, $H\rightarrow H-i\epsilon(\vecr)$.
We adopt a function form of $\epsilon(r)$ being linearly dependent on
the radial coordinate $r$.
The conditions and limitations on $\epsilon(\vecr)$ are discussed in
a number of works (see Ref.~\cite{SM92} and references therein).

In this work, we use ABC in the time-dependent Hartree-Fock calculations
on a 3D coordinate grid.
In the real-time calculations, the linear response is computed by
applying an impulsive external field to the Hartree-Fock (HF) ground state,
then calculating the expectation values of some observables as a function
of time, and then Fourier transforming to get the energy response
\cite{YB96,NY01-1,NY02-2}.
Since all frequencies are contained in the initial perturbation,
the entire energy response can be calculated with a single time evolution.

We use the Skyrme energy functional of Ref.~\cite{BFH87} with
the SGII parameter set \cite{GS81}.
For calculations of the HF ground state,
the imaginary-time method of Ref.~\cite{Dav80} is utilized.
For the time evolution of the TDHF state,
to which the time-odd components in
the energy functional also contribute,
we follow the prescription in Ref.~\cite{FKW78}.
The model space is a sphere whose radius is 22 fm.
The absorbing potential,$-i\epsilon(r)$, is zero in a region of $r<10$ fm,
while it is non-zero at $r>10$ fm.
The TDHF single-particle wave functions are discretized on a rectangular mesh
in a 3D real space.
In order to reduce number of mesh points outside the interaction region,
we employ curvilinear coordinates which are mapped by a change of coordinates
to an adaptive mesh in Cartesian coordinates \cite{MZK97}.
More details will be published in our forthcoming paper \cite{NY02-3}.

In this paper, we discuss an application of ABC to the isovector
giant dipole resonance (GDR) in $^{24}$Mg.
The density distribution of the ground state has a prolate shape.
In Table~\ref{tab:spe}, calculated single-particle energies of
neutrons and protons are listed.
The deformation lifts the $(2j+1)$-fold but not the Kramers degeneracy.
The energies of the highest occupied single-particle orbitals
for neutrons and protons
are $-14.4$ MeV and $-9.7$ MeV, respectively.
Since the height of Coulomb barrier for protons is $4\sim 5$ MeV,
the nucleonic continuum plays an effective role when the nucleus gets
excitation energy larger than $14\sim 15$ MeV.

We take an initial external field as
\begin{equation}
{\bf V}_{\rm ext}(t) = k\vecr
  \left\{\frac{1}{2}\left( 1-\tau_z\right)e-\frac{Ze}{A}\right\}
  \delta(t) ,
\label{Vext}
\end{equation}
where $k$ should be small enough to validate
the linear response approximation.
Of course, after the Fourier transform,
the calculated $E1$ strength should not
depend on the magnitude of $k$.
In this calculation, we choose $k=0.001$.

\begin{figure}
\resizebox{0.5\textwidth}{!}{%
  \includegraphics{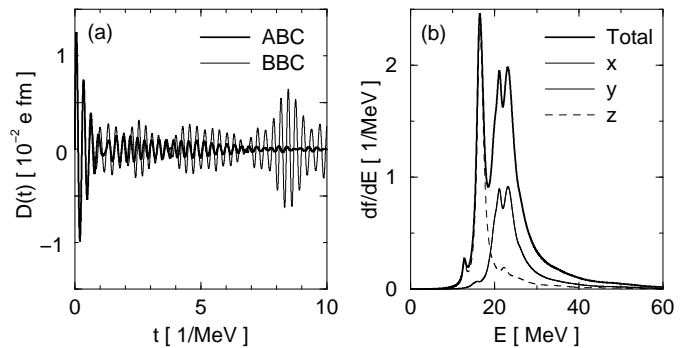}
}
\caption{(a) Calculated $E1$ moment as a function of time, $D_x(t)$,
for $^{24}$Mg with ABC (thick line) and
with BBC (Box boundary condition: thin line).
The direction of the initial external dipole field is taken to be
perpendicular to the symmetry axis ($z$-axis).
(b) $E1$ oscillator strength calculated by Fourier transforming ${\bf D}(t)$.
}
\label{fig:Mg}       % Give a unique label
\end{figure}

Fig.~\ref{fig:Mg} (a) shows that the time evolution of $E1$ dipole
moment initiated by the impulsive perturbation of Eq.~(\ref{Vext}),
$(V_{\rm ext})_x$.
We perform two kinds of TDHF simulation with the same initial perturbation;
TDHF with the ABC described above, and
the one with the box (vanishing) boundary condition at $r=10$ fm.
Time evolution is carried out up to $T=10\ \hbar$/MeV.
The total period of time evolution, $T$, is related with
the energy resolution after the Fourier transform, $\Delta E$,
by $\Delta E=2\pi\hbar/T$.
In the box calculation, the dipole oscillation damps in a short period of
time.
However, we find a beating structure in a long period.
The beat comes from reflection of outgoing waves during the time evolution
of the TDHF state.
On the other hand, the ABC calculation does not indicate
any prominent beating pattern.
Since the difference becomes significant at $t\ge 1\ \hbar$/MeV,
the box calculation can provide a reliable result only when the strength
function is averaged over an energy width of 6 MeV.

The Fourier transform of the dipole moment,
$D_x(t)$, $D_y(t)$, and $D_z(t)$, calculated with ABC
leads to $E1$ oscillator strengths of $^{24}$Mg shown in Fig.~\ref{fig:Mg} (b).
Here, we use a smoothing parameter of $\Gamma=1$ MeV.
The calculation shows a prominent double-peak structure which corresponds to
the deformation splitting of the GDR.
The distribution of the oscillator strength qualitatively agrees
with the photoneutron cross section \cite{BF75}.
A high energy tail of the oscillator strength (at $E>25$ MeV)
cannot be reproduced in the calculation with the box boundary condition.
It turns out that the box calculation
significantly changes the $E1$ strength at $E\ge 23$ MeV,
compared to the ABC results of Fig.~\ref{fig:Mg} (b).
In other words, the damping effect caused by fast emissions of a nucleon
is important for $E\ge 23$ MeV.

In summary, the isovector GDR in the continuum of deformed nuclei
is investigated with the linear response theory.
The single-nucleon continuum is taken into account by the ABC.
The real-time TDHF is an efficient method to calculate the energy response
over a wide range of energy.
We consider that the present method, {\it TDHF combined with ABC},
is promising and we are studying excited states and
responses in nuclei near drip lines \cite{NY02-1,NY02-3}.

% table
\begin{table}
\caption{Occupied single-particle energies for $^{24}$Mg calculated with
SGII.  Each level has a two-fold degeneracy.}
\label{tab:spe}       % Give a unique label
\begin{center}
\begin{tabular}{cc}
\hline\noalign{\smallskip}
 neutrons [ MeV ] & protons [ MeV ]  \\
\noalign{\smallskip}\hline\noalign{\smallskip}
 -40.6 & -35.4 \\
 -29.8 & -24.9 \\
 -24.4 & -19.5 \\
 -19.8 & -14.8 \\
 -17.4 & -12.6 \\
 -14.4 & -9.70 \\
\noalign{\smallskip}\hline
\end{tabular}
\end{center}
\end{table}

% figure

\end{document}